\documentclass[aps,preprint,epsfig,rotate]{revtex4}
\usepackage{graphicx}
\usepackage{bm}
\usepackage{epsfig}

\input epsf
\begin{document}
\title{Spectrum of secondary electrons emitted during the nuclear $\beta^{-}$-decay of the tritium atom.}

\author{Alexei M. Frolov}
\email[E--mail address: ]{afrolov@uwo.ca}

\affiliation{Department of Applied Mathematics \\
 University of Western Ontario, London, Ontario N6H 5B7, Canada}

\date{\today}

\begin{abstract}
Ionization of the final ${}^{3}$He$^{+}$ ion during the nuclear $\beta^{-}$-decay of the tritium atom is discussed. The velocity spectrum of the emitted 
secondary electrons is derived in the explicit form. Our method allows to determine the relative and absolute probabilities of formation of the final 
states in few-electron atoms which include `free' secondary electrons moving with different velocities. 

\noindent 
PACS number(s): 23.40.-s, 12.20.Ds and 14.60.Cd


\end{abstract}

\maketitle
\newpage

In this study we investigate the velocity spectrum and a few other properties of the `free' electrons emitted during the nuclear $\beta^{-}-$decay of atomic nuclei. Our goal 
is to derive the closed analytical formula for the spectral function of such secondary electrons and determine the conditional and total probabilities of their emission. As 
is well known the velocity of the fast $\beta^{-}$-electrons ($v_{\beta}$) emitted during the nuclear $\beta^{-}-$decay of atomic nuclei are significantly larger than typical 
velocities of bound atomic electrons $v_a$. In light atoms we have $v_{\beta} \ge 100 v_a - 1000 v_a$. The inequality $v_{\beta} \gg v_a$ allows one to apply the sudden 
approximation and analyze the nuclear $\beta^{-}$-decay in light atoms by calculating the overlaps of the non-relativistic atomic wave functions. The sudden approximation is 
based on the fact that the wave function of incident atomic system does not change during the fast process, i.e. its amplitude and phase do not change. This means that 
electron density distribution in the incident atom does not change during $\beta^{-}$-decay of its nucleus (see discussions in \cite{LLQ} and \cite{MigK}). This allows one to 
determine all probabilities of the bound-bound and bound-free transitions, i.e. the $p_{bb}$ and $p_{bb}$ values. By the transition we mean the actual transition during the 
nuclear $\beta^{-}$-decay from one bound state in the incident atom into the final (bound, or unbound) state in the final ion. 

To avoid a very general discussion with use of very complex notations for atomic terms, let us consider the nuclear $\beta^{-}$-decay of the tritium atom which has only one 
bound electron. Moreover, for simplicity in this study we restrict our analysis to the case when the incident tritium atom was in its ground $1^2s-$state (before 
$\beta^{-}$-decay). The nuclear $\beta^{-}$-decay of the tritum atom proceeds in one of the following ways (see, e.g., \cite{Blat}, \cite{Fro05})
\begin{eqnarray}
  & & {}^{3}{\rm H} \rightarrow {}^{3}{\rm He}^{+} + e^{-}(\beta) + \overline{\nu} \label{eq0} \\
  & & {}^{3}{\rm H} \rightarrow {}^{3}{\rm He}^{2+} + e^{-} + e^{-}(\beta) + \overline{\nu} \label{eq1}
\end{eqnarray}
where the notation $e^{-}(\beta)$ designates the fast $\beta^{-}-$electron, $\overline{\nu}$ denotes the electron's anti-neutrino, while the notation $e^{-}$ stands for the 
secondary (or slow) electron formed in the unbound spectrum during the $\beta^{-}-$decay of the tritum atom. Below, the electric charge of incident nucleus ($Q$) is designated 
by the notation $Q_1$, while the electric charge of the final nucleus is denoted by the notation $Q_2 (= Q + 1)$. Numerical computations of the probabilities of the bound-bound 
transitions for the process, Eq.(\ref{eq0}), are performed since earlier papers by Migdal (references can be found, e.g., in \cite{MigK}). In general, such calculations are 
simple and straightforward. Currently, the overall accuracy of numerical computations of the bound-bound probabilities is relatively high (see, e.g., \cite{Fro05} - \cite{Our1}). 
For instance, by using the explicit formulas for the one-electron wave functions of the ${}^{3}$H atom and ${}^{3}$He$^{+}$ ion we have found that the total probability of the 
bound-bound transitions for the process, Eq.(\ref{eq0}), equals $P_{bb}$ = 0.97372735(10) (see Table I). The difference between unity and $P_{bb}$ value is the total probability 
of the bound-free transitions $P_{bf}$ = 1 - $P_{bb} \approx$ 0.02627265(10) during the nuclear $\beta^{-}$-decay of the tritium atom with infinitely heavy nucleus. In many 
experiments it is important to know partial probabilities $p_{bf}({\bf p})$ of the bound-free transitions, rather than the $P_{bf}$ value. In earlier papers this problem has not 
been solved. Therefore, at this moment we do not know the velocity/momentum spectra of the secondary electrons emitted during nuclear $\beta^{-}$-decay in atoms. 

In this study we consider the general theory of the bound-free transitions and derive the formulas which allow us to evaluate the partial probabilities $p_{bf}({\bf p})$ of such
transitions. We also obtain the explicit formulas to represent the velocity/momentum spectrum of the secondary electrons. For the tritium atom such a spectral function is 
relatively simple and unique, but in few-electron atoms/ions the shape and other parameters of such a spectral function of secondary electrons depends upon electron-electron 
correlations in the incident atom. 

In sudden approximation the final state probability of the process, Eq.(\ref{eq1}), equals to the overlap integral of the wave functions of the incident tritium atom ${}^{3}$H 
and wave function of the final ${}^{3}$He$^{2+}$ ion multiplied by the wave function of the outgoing (or `free') electron which has a certain momentum ${\bf p}$. The direction 
of the momentum ${\bf p}$ in space coincides with the direction of motion/propagation of the actual free electron that is observed in experiments. Moreover, at large 
electron-nucleus distances each of these free-electron wave functions must be a linear combination of a plane wave and incoming spherical wave. Functions with such an asymptotic 
at large $r$ take the form \cite{Maxim} (see also \$ 136 in \cite{LLQ})
\begin{eqnarray}
 \phi_{p}(r, {\bf n}_p \cdot {\bf n}_r) = N_f \exp(\frac{\pi}{2} \zeta) \Gamma(1 + \imath \zeta) \cdot {}_1F_1\Bigl(-\imath \zeta, 1, -\imath ({\bf p} \cdot {\bf r} - p r)\Bigr) 
 \exp[\imath ({\bf p} \cdot {\bf r})] \label{Cwave} 
\end{eqnarray}
where $N_f = \frac{1 - \exp(-2 \pi \zeta)}{\sqrt{2 \pi \zeta}}$ is the normalization constant, ${}_1F_1(a, b; z)$ is the confluent hypergeometric function and $\zeta = 
\frac{Q_2}{a_0 p} = \frac{\alpha Q_2}{\gamma v}$, where $a_0 = \frac{\hbar^2}{m_e e^2}$ is the Bohr radius, $\alpha = \frac{e^2}{\hbar c}$ is the fine structure constant and $\gamma$ 
is the Lorentz $\gamma-$factor of the moving electron. The notations $p$ and $v$ stand for the absolute values of the momentum and velocity of the outgoing (or `free') electron. Also 
in this equation the two unit vectors ${\bf n}_p$ and ${\bf n}_r$ are defined as follows ${\bf n}_p = \frac{{\bf p}}{p}$ and ${\bf n}_r = \frac{{\bf r}}{r}$. 

The ground $1^2s-$state wave function of the one-electron, hydrogen-like atom/ion is $\frac{\eta \sqrt{\eta}}{\sqrt{\pi}} \exp(-\eta r)$, where $\eta = \frac{Q}{a_0}$ (in atomic units 
where $\hbar = 1, m_e = 1$ and $e = 1$). Below, the following system of notations is applied for the $\beta^{-}$ decaying tritium atom: $Q_1 = Q = 1, \eta = \frac{Q_1}{a_0}$, while 
for the final helium ion He$^{+}$ we chose $Q_2 = Q + 1(= 2)$ and $\zeta = \frac{Q_2}{a_0 p} = \frac{\alpha Q_2}{\gamma v}$. The probability amplitude equals the overlap integral 
between the $\frac{\eta \sqrt{\eta}}{\sqrt{\pi}} \exp(-\eta r)$ function and the $N_f \phi_{kl}(r, {\bf n}_p \cdot {\bf n}_r)$ function, Eq.(\ref{Cwave}). This leads to the following 
formula for the overlap integral: 
\begin{eqnarray}
  & & I_2(\eta) = 4 \pi \int \exp[\imath ({\bf p} \cdot {\bf r} - \eta r)] {}_1F_1\Bigl(-\imath \zeta, 1, -\imath ({\bf p} \cdot {\bf r} - p r)\Bigr) r^{2} dr \nonumber \\
  &=& -\frac{\partial I_1(\eta)}{\partial \eta} = 8 \pi \frac{\eta + \zeta p}{(\eta^2 + p^2)^2} \exp\Bigl[-2 \zeta \arctan\Bigl(\frac{\eta}{p}\Bigr)\Bigr]  \label{Max5}
\end{eqnarray}
where analogous integral $I_1(\eta)$ has been determined (analytically) in \cite{Maxim}. The $I_2(\eta)$ integral, Eq.(\ref{Max5}) (with the additional normalization factors $N_f$ and 
$N_{{\rm H}}$) determines the probability amplitude of the electron ionization of the helium-3 atom during the nuclear $\beta^{\pm}$ decay of the incident hydrogen/tritium atom, 
which was originally in its ground $1^2s$-state. The momentum of the `free' electron is ${\bf p}$ and $p = \mid {\bf p} \mid$ is its absolute value. If we want to determine the final 
state probabilities of atomic ionization during nuclear $\beta^{\pm}$ decay of the hydrogen/tritium atom from the excited $s-$states, then higher derivatives from the $I_1(\eta)$ 
integral \cite{Maxim} in respect with the $\eta$ variables are needed. Finally, for the $\beta^{-}$ decay from the ground $1^{2}s-$state of the ${}^{3}$H atom one finds the following 
formula for the probability amplitude ${\cal A}$
\begin{eqnarray}
  {\cal A} = 8 \pi N_H N_f \cdot \frac{\eta \Bigl(\frac{Q_2}{Q_1} + 1\Bigr)}{(\eta^2 + p^2)^2} \cdot \exp\Bigl[-2 \Bigl(\frac{Q_2 \eta}{Q_1 p}\Bigr) 
  \arctan\Bigl(\frac{Q_2 \eta}{Q_1 p}\Bigr)\Bigr] \label{Max54}
\end{eqnarray}
where $N_{{\rm H}} = \sqrt{\frac{\eta^3}{\pi a^{3}_{0}}}$ is the normalization constant of the hydrogen-atom wave function, while $N_f = \sqrt{\frac{1 - \exp(-2 \pi \zeta)}{2 \pi 
\zeta}}$ is the normalization constant of the wave function which represents the `free' electron. The expression for the infinitely small final state probability ($\Delta 
P_{i \rightarrow f} \simeq \mid {\cal A} \mid^2$) takes the form
\begin{eqnarray}
 & & \Delta P_{i \rightarrow f} = \mid {\cal A} \mid^2 p^2 \Delta p = \frac{32 \eta^3}{\zeta} \cdot \Bigl[1 - \exp\Bigl(-2 \pi \frac{Q_2 \eta}{Q_1 p}\Bigr)\Bigr] \cdot 
 \frac{p^2 \eta^2 \Bigl(\frac{Q_2}{Q_1} + 1\Bigr)^2}{(\eta^2 + p^2)^4} \times \nonumber \\
 & & \exp\Bigl[-4 \Bigl(\frac{Q_2 \eta}{Q_1 p}\Bigr) \arctan\Bigl(\frac{Q_2 \eta}{Q_1 p}\Bigr)\Bigr] \Delta p  \label{Max56}
\end{eqnarray} 
To produce the final expression which can be used in calculations we have to replace here the variables $\eta$ and $\zeta$ by the following expressions $\eta = \frac{Q_1}{a_0}, 
\frac{\eta}{p} = \frac{\alpha Q_1}{\gamma v}$ and $\zeta = \frac{Q_2 \eta}{Q_1 p} = \frac{\alpha Q_2}{\gamma v}$, where $Q_1(= Q)$ is the electric charge of the incident `bare' nucleus 
(or central positively charged ion) and $a_0 = \frac{\hbar^2}{m_e e^2}$ is the Bohr radius. In atomic units, where $\hbar = 1, e = 1$ and $m_e = 1$, the Bohr radius equals unity and the 
ratio $\frac{\eta}{p}$ equals to the ratio $\frac{\alpha Q_1}{\gamma v}$ (since $m_e = 1)$, where $\alpha = \frac{\hbar^2}{m_e e^2}$ is the fine structure constant and $v = \mid {\bf v} 
\mid$ is the absolute value of the electron's velocity (expressed in atomic units). The factor $\gamma = \frac{1}{\sqrt{1 - \frac{v^2}{c^2}}} = \frac{1}{\sqrt{1 - \alpha^2 v^2}}$ is the 
Lorentz $\gamma-$factor of the moving electron. Numerically in atomic units the electron's velocity $v$ cannot exceed the value of $c = \alpha^{-1} (\approx$ 137 in $a.u.$).  

This allows one to obtain the following expression for the $v-$spectral function of the secondary electron emitted in the process, Eq.(\ref{eq1}) (or $v-$spectrum, for short):
\begin{eqnarray}
 & & S_e(v; Q) = \frac{32 Q_1}{{\cal S}(Q) \alpha Q_2} \cdot \Bigl[1 - \exp\Bigl(-2 \pi \frac{Q_2 \alpha}{\gamma v}\Bigr)\Bigr] \cdot \frac{(Q^{2}_{1} + Q^{2}_{2})^2 \gamma^{4} 
  v^{3}}{(Q^{2}_1 + \gamma^2 v^2)^4} \times \nonumber \\
 & & \exp\Bigl[-4 \Bigl(\frac{\alpha Q_2}{\gamma v}\Bigr) \arctan\Bigl(\frac{\alpha Q_2}{\gamma v}\Bigr)\Bigr] \label{Max567}
\end{eqnarray} 
where the normalization constant ${\cal S}(Q)$ must be chosen from the condition that integral of $S_e(v; Q)$ over $v$ from 0 to $v_{max}$ must be equal unity. Numerical value of this 
constant can be found (for each pair $Q_1, Q_2$, where $Q_1 = Q$ and $Q_2 = Q + 1$), by using methods of numerical integration \cite{Num}. In actual applications to few- and many-electron 
atoms we have to take into account the known fact that all bound atomic electrons are non-relativistic particles. The corresponding velocities of internal electrons $v$ are substantially 
less than $\frac{c}{4}$. For light atoms such `atomic' velocities do not exceed the value $\approx \frac{c}{5}$. Moreover, in our calculations of the overlap integral both non-relativistic 
wave functions have been applied. It follows from here that the non-relativistic approximation is more appropriate to describe properties of secondary electrons from Eq.(\ref{eq1}). This 
means that in Eq.(\ref{Max567}) we have to assume that $\gamma = 1$, i.e. Eq.(\ref{Max567}) takes the form
\begin{eqnarray}
 & & S_e(v; Q) = \frac{32 Q_1}{{\cal S}(Q) \alpha Q_2} \cdot \Bigl[1 - \exp\Bigl(-2 \pi \frac{Q_2 \alpha}{v}\Bigr)\Bigr] \cdot \frac{(Q^{2}_{1} + Q^{2}_{2})^2 v^{3}}{(Q^{2}_1 + v^2)^4} 
 \times \nonumber \\
 & & \exp\Bigl[-4 \Bigl(\frac{\alpha Q_2}{v}\Bigr) \arctan\Bigl(\frac{\alpha Q_2}{v}\Bigr)\Bigr] \label{Max568}
\end{eqnarray} 
where $v$ varies between 0 and $v_{max} = 50 \alpha c$ (this value of $v_{max}$ can be used for any light atom and/or ion). However, in this study we apply the spectral function,
Eq.(\ref{Max567}).  
 
By using the formula, Eq.(\ref{Max567}), for the $\beta^{-}-$decay of the tritum atom with an infinitely heavy nucleus we have found that ${\cal S}(Q) \approx$ 196.611833628395. As 
expected this formula contains only the absolute values of free-electron velocity $v$ (or momentum $p$) and electric charge of the incidnet atomic nucleus $Q$. The velocity of the fast 
$\beta^{-}-$electron is not included in this formula. This is a direct consequence of the sudden approximation which has been used to derive the formulas, Eq.(\ref{Max567}) and 
Eq.(\ref{Max568}). In general, by using the known $v$-spectral function we can evaluate the probability $p(v)$ to observe a secondary electron which moves with the velocity $v$, where $v 
\ll c (= \alpha^{-1}$ in $a.u.$). In general, the integral from the spectral function $S_e(v; Q)$ between the $v_1$ and $v_2$ values ($v_2 > v_1$) gives one the probability $P(v_1;v_2)$ 
to detect the `free' electron with the velocity bounded between the $v_1$ and $v_2$ values. This probability is normalized to all unbound spectra of the final ion. All states of the 
discrete spectrum are ignored during this procedure. In many actual cases, however,  it is important to determine the absolute probability $\overline{P}(v_1;v_2)$ of the bound-free 
transitions during nuclear nuclear $\beta^{-}$ decay, i.e. in those cases when the states of of discrete spectrum are included in calculations of probabilities. To obtain this value we 
have to apply the $P_{bf}$ (or $P_{bf}$) quantity which has beed evaluated above. Then, we can write the following formula for the conditional probability
\begin{eqnarray}
   \overline{P}(v_1;v_2) = P_{bf} P(v_1;v_2) = (1 - P_{bb}) P(v_1;v_2) \label{prob}
\end{eqnarray} 
Numerical values of such probabilities $P(v_1;v_2)$ computed with the unity step ($v_2 = v_1 + 1$) can be found in Table II. Note that for the process, Eq.(\ref{eq1}), the most important 
velocities $v$ are located between $v_{min} \approx$ 0.4 and $v_{max} \approx$ 3.4. Numerical values of the final state probabilities determined for the different velocity intervals 
$[ v_1, v_2]$ can be found in Table II, where the formula, Eq.(\ref{Max567}), is used. In this paper we can present a very short vesrion of this Table. For light atoms, the probabilities 
determined with the use of both spectral functions, Eqs.(\ref{Max567}) and (\ref{Max568}) are always very close to each other. This follows from internal structure of thee functions which 
contains an exponential `cutt-off' factor, which essentially removes all large energies. By using the spectral functions, Eqs.(\ref{Max567}) - (\ref{Max568}), we can determine all 
bound-free transition probabilities for the $\beta^{-}$ decay in the tritium atom, Eq.(\ref{eq1}).
  
Formulas derived in this study allows one to determine all final state probabilities for the $\beta^{-}$-decaying tritium atom. We have develop an approach which succssefully works to 
determine the final state probabilities of the bound-free transitions during the nuclear $\beta^{-}$ decay of the one-electron tritium atom. Our wave functions of the final electron 
represent the actual electron which moves `free' in the field of the final He$^{+}$ nucleus. The same approach can be used to derive the explicit formulas for the final state probabilities 
and velocity/momentum spectra of the secondary electrons which arise during nuclear $\beta^{-}$ decay of an arbitrary few-electron atom. Preliminary investigations of such few-electron 
atoms incidate clearly that spectra of secondary electrons have different forms for different few-electron atoms/ions. The most important atomic factor which substantially changes the 
actual spectra of secondary electrons emitted dutring nuclear $\beta^{-}$ decay of few-electron atoms is related to the electron-electron correlations in the incident atom/ion.

\begin{table}[tbp]
   \caption{Convergence of the total probabilities $P_{bb}$ of the bound-bound transitions during the nuclear $\beta^{-}$ decay of the 
            tritium atom with an infinitely heavy nucleus. $N$ is the total number of hydrogen $ns$-states used in calculations, 
            $n$ is the principal quantum number, while the notation $s$ corresponds to the electron angular momentum $\ell$ equals zero.}
     \begin{center}
     \begin{tabular}{| c | c | c | c |}
      \hline\hline
 $N$  &  $P_{bb}$ & $N$  &  $P_{bb}$ \\ 
         \hline
   100 & 0.97371867838323 &  500 & 0.97372684019166 \\
  1000 & 0.97372709722987 & 1500 & 0.97372714487989 \\
  1600 & 0.97372714949736 & 1700 & 0.97372715332442 \\
         \hline \hline
  \end{tabular}
  \end{center}
  \end{table}
\begin{table}[tbp]
   \caption{Probabilities of the bound-free transitions $p_{bf}(v_1,v_2)$ during the nuclear $\beta^{-}$ decay of the 
            tritium atom with an infinitely heavy nucleus. Calculations are performed with the use of the formula, 
            Eq.(\ref{Max567}), where $0 \le v \le \alpha^{-1}$. To obtain the absolute final state probabilities these 
            values must be multiplied by the additional factor $P_{bf} \approx 0.02627265(10)$.}
     \begin{center}
     \scalebox{0.80}{%
     \begin{tabular}{| c | c | c | c | c | c | c | c | c |}
      \hline\hline
 $v_1$ & $v_2$ & $p_{bf}(v_1,v_2)$ &  $v_1$ & $v_2$ & $p_{bf}(v_1,v_2)$ &  $v_1$ & $v_2$ & $p_{bf}(v_1,v_2)$ \\ 
         \hline
  0.1 & 0.2 & 0.1887486938E-02 & 3.3 & 3.4 & 0.7074717265E-02 & 11.4 & 11.5 & 0.1329840115E-04 \\
  0.4 & 0.5 & 0.1917130375E-01 & 3.6 & 3.7 & 0.5015703351E-02 & 12.6 & 12.7 & 0.7446739450E-05 \\
  0.5 & 0.6 & 0.2681487042E-01 & 3.7 & 3.8 & 0.4480407262E-02 & 13.0 & 13.1 & 0.6206926650E-05 \\
  1.0 & 1.1 & 0.5254300663E-01 & 4.4 & 4.5 & 0.2096616982E-02 & 15.0 & 15.1 & 0.2680741819E-05 \\
                     \hline                        
  1.4 & 1.5 & 0.4992373809E-01 & 5.0 & 5.1 & 0.1262893513E-02 & 17.0 & 17.1 & 0.1277214108E-05 \\
  1.5 & 1.6 & 0.4723742821E-01 & 5.2 & 5.3 & 0.9451119318E-03 & 18.0 & 18.1 & 0.9084928397E-06 \\
  1.6 & 1.7 & 0.4416838598E-01 & 5.4 & 5.5 & 0.7835661935E-03 & 19.0 & 19.1 & 0.6574626020E-06 \\
  2.0 & 2.1 & 0.3102588126E-01 & 6.7 & 6.8 & 0.2570619146E-03 & 23.0 & 23.1 & 0.2077424148E-06 \\
        \hline
  2.5 & 2.6 & 0.1798174511E-01 & 8.4  &  8.5 & 0.750897540E-04 & 35.0 & 35.1 & 0.1552228745E-07 \\
  2.7 & 2.8 & 0.1426319226E-01 & 9.2  &  9.3 & 0.451506651E-04 & 45.0 & 45.1 & 0.3080476662E-08 \\
  3.0 & 3.1 & 0.1003778633E-01 & 10.4 & 10.5 & 0.225258543E-04 & 60.0 & 60.1 & 0.4305802692E-09 \\
  3.2 & 3.3 & 0.7945901258E-02 & 11.0 & 11.1 & 0.163344103E-04 & 75.0 & 75.1 & 0.7910055696E-10 \\
        \hline \hline
  \end{tabular}}
  \end{center}
  \end{table}
\end{document}